\documentclass[english]{article}
\usepackage[T1]{fontenc}
\usepackage[latin9]{inputenc}
\usepackage{amsmath}
\usepackage{amssymb}
\usepackage{graphicx}

\makeatletter

\providecommand{\tabularnewline}{\\}

\newcommand{\lyxaddress}[1]{
\par {\raggedright #1
\vspace{1.4em}
\noindent\par}
}

\makeatother

\usepackage{babel}
\begin{document}

\title{\textbf{Observation of a Fast Evolution in a Parity-time-symmetric
System}}

\author{Chao Zheng$^{1}$, Liang Hao$^{1}$ and Gui Lu Long$^{1,2}$}

\maketitle

\lyxaddress{}

\lyxaddress{\begin{center}
\textit{$^{1}$State Key Laboratory of Low-dimensional Quantum Physics
and Department of Physics, Tsinghua University, Beijing 100084, P.
R. China}\\
 \textit{ $^{2}$Tsinghua National Laboratory for Information Science
and Technology,}\\
 \textit{ Beijing 100084, P. R. China }
\par\end{center}}

\begin{flushleft}
In the parity-time-symmetric (PT-symmetric) Hamiltonian theory, the
optimal evolution time can be reduced drastically and can even be
zero. In this letter, we report our experimental simulation of the
fast evolution of a PT-symmetric Hamiltonian in a nuclear magnetic
resonance quantum system. The experimental results demonstrate that
the PT-symmetric Hamiltonian system can indeed evolve much faster
than the quantum system, and the evolution time can be arbitrarily close
to zero.
\par\end{flushleft}

\begin{center}
\textbf{Keywords: parity-time-symmetric Hamiltonian; non-Hermiticity;
nuclear magnetic resonance; brachistochrone; duality quantum computation.}
\par\end{center}

\begin{center}
\begin{tabular}{cccccccccccccccccccccccccc}
\hline
 &  &  &  &  &  &  &  &  &  &  &  &  &  &  &  &  &  &  &  &  &  &  &  &  & \tabularnewline
\end{tabular}
\par\end{center}

\begin{center}
\textbf{\large 1. Introduction}{\large{} }
\par\end{center}{\large \par}

The brachistochrone problem, i.e., the minimum time evolution between
two states, is an interesting and important problem. In quantum mechanics,
the brachistochrone between two states is bounded by the maximum difference
of the eigenvalues of the Hamiltonian \cite{qbrac1,qbrac2,qbrac3,qbrac4,qbrac5,qbrac5-1,qbrac7,qbrac8}.
Brachistochrone has important applications. For instance, the time-optimal
approach to the quantum algorithmic complexity has attracted much
interest recently \cite{optime1,optime2}.


The Hermiticity requirement of a Hamiltonian guarantees that its eigenvalues
are real. It also implies that the evolutionary operator $e^{-\frac{i}{\hbar}Ht}$
is unitary. However, Hermiticity is a sufficient condition but not
necessary for real eigenvalues, and the entire spectrum of a wide
class of non-Hermitian Hamiltonians can also be real. Among these
Hamiltonians \cite{th1} is a class that is PT symmetric. The PT-symmetric
Hamiltonian has been investigated intensively in recent years, both
in theory \cite{th1,th2,th3,th3-1,th3-2,th4,th4-1,ptbrac,th7,th8,th9,th11-1,th12,equal,th12-1,th15,th18,th19,th20,th21,th22}
and in experiments \cite{exp1,exp2,exp3,exp4,exp5}. PT symmetries
have been experimentally observed in table-top optical systems \cite{exp1,exp2,exp3,exp4}
and in spin-polarized Rb atoms \cite{exp5}.

The novel character of PT-symmetric Hamiltonians brings about many
new features and may lead to interesting applications. Faster than Hermitian quantum mechanics
evolution is one of these important aspects \cite{ptbrac}. In this
article, we design and carry out an experiment that simulates the time
evolution of a PT-symmetric Hamiltonian with a nuclear magnetic resonance
(NMR) quantum system. We build a system in Hilbert space that admits
both unitary and non-unitary evolution, and observe the time evolution
of a PT-symmetric Hamiltonian
The
experimental result shows that the minimal evolutionary time in a
PT-symmetric system is faster than that in the Hermitian case, and
can be arbitrarily close to zero.


$ $

\begin{center}
\textbf{\large 2. Theoretical Frame}{\large{} }
\par\end{center}{\large \par}

A simple PT-symmetric non-Hermitian Hamiltonian for a two-level system
is
\begin{equation}
H=\left(\begin{array}{cc}
se^{i\alpha} & s\\
s & se^{-i\alpha}
\end{array}\right).\label{e2}
\end{equation}
 According to Ref. \cite{ptbrac}, the largest and smallest eigenvalues
are $E_{\pm}=2s\cos\alpha$ and $0$, respectively. The difference
between them is
\begin{equation}
\omega=E_{+}-E_{-}=2s\cos\alpha.\label{e3}
\end{equation}
 Under the influence of $e^{-\frac{i}{\hbar}Ht}$, the PT-symmetric
system that is initially in $|0\rangle=(1\;\;0)^{T}$ will go to \cite{ptbrac}
\begin{equation}
e^{-\frac{i}{\hbar}Ht}\left(\begin{array}{c}
1\\
0
\end{array}\right)=\frac{e^{-\frac{i}{\hbar}ts\cos\alpha}}{\cos\alpha}\left(\begin{array}{c}
\cos\left(\frac{\omega t}{2\hbar}-\alpha\right)\\
-i\,\sin\left(\frac{\omega t}{2\hbar}\right)
\end{array}\right).\label{e4}
\end{equation}
 It takes the time
\begin{equation}
\tau=\frac{2\hbar}{\omega}\left(\alpha+\frac{\pi}{2}\right)\label{e5}
\end{equation}
 to evolve to state $|1\rangle=(0\;\;1)^{T}$. When $\alpha\rightarrow-\pi/2$,
it approaches zero, which is an impossible task for a regular Hermitian
Hamiltonian.

For comparison, the equivalent Hermitian Hamiltonian, $H_{0}$, was
calculated \cite{ptbrac}
\begin{equation}
{H_{0}}=\left(\begin{array}{cc}
s\,\cos\alpha & s\,\cos\alpha\\
s\,\cos\alpha & s\,\cos\alpha
\end{array}\right),\label{e6}
\end{equation}
 the eigenvalues of which are $E_{\pm}=2s\cos\alpha$ and $0$, respectively.
Here, $E_{+}-E_{-}=\omega$, which is exactly the same as that in
the PT-symmetric case.

The evolution under this equivalent Hermitian Hamiltonian is given
by
\begin{equation}
e^{-\frac{i}{\hbar}{H_{0}}{t}}\left(\begin{array}{c}
1\\
0
\end{array}\right)=e^{-\frac{i}{\hbar}{t}s\cos\alpha}\left(\begin{array}{c}
\cos\left(\frac{s\,\cos\alpha}{\hbar}{t}\right)\\
-i\,\sin\left(\frac{s\,\cos\alpha}{\hbar}{t}\right)
\end{array}\right),\label{e7}
\end{equation}
 and the time it takes to evolve to the final state $|1\rangle$ is
\begin{equation}
{\tau_{0}}=\frac{\pi\hbar}{\omega},\label{e8}
\end{equation}
 which is constant for a fixed $\omega$.

$ $

\begin{center}
\textbf{\large 3. Construction of a PT-symmetric Hamiltonian System}{\large{} }
\par\end{center}{\large \par}

\begin{figure}
\centering{}\includegraphics[scale=0.36]{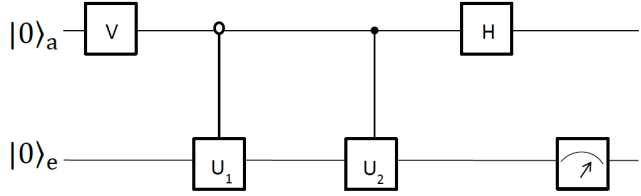} \caption{ Quantum circuit for a PT-symmetric system. Operations from left to
right are a single-qubit operation $V$, $C_{0-U1}$ (a 0-controlled
$U_{1}$), $C_{1-U2}$ (a 1-controlled $U_{2}$), and a Hadamard operation
$H$. When the ancillary qubit is in $|0\rangle_{a}$, the final state
of work qubit is $e^{-\frac{i}{\hbar}{H}{t}}|0\rangle_{e}$. \label{f1}}
\end{figure}

We now construct a system with a PT-symmetric Hamiltonian Eq. (\ref{e2})
and simulate its time evolution. The vital part of the simulation
is achieved using the idea of duality quantum computing \cite{duality,duality2}.  Duality quantum computing can be achieved by using an ancilla qubit using a conventional quantum computer  \cite{duality,duality2}.
The principle to simulate a duality quantum computing is shown in a quantum circuit in Fig. \ref{f1}.
Simulating the non-unitary evolution of a PT-symmetric quantum system requires the determination of the explicit forms of the operators in Fig. \ref{f1}.

For a conventional quantum computer, the idea to use an extended
space consisted of a system and an ancilla for simulating a non-unitary
evolution in a PT-symmetric system by a unitary evolution in an extended space was proposed
in Ref. \cite{ancilla}. This is similar to the our scheme used in the experiment in this work.

The system we use contains two qubits: the work qubit $e$ and the ancillary
qubit $a$. A qubit is a two-level quantum system that is the building
block in quantum information processing. The input 2-qubit state on
the leftmost is $|0\rangle_{a}|0\rangle_{e}$. We then perform the
$V$ unitary operation,
\begin{eqnarray}
V & = & \left(\begin{array}{cc}
\cos\phi_{V} & -\sin\phi_{V}\\
\sin\phi_{V} & \cos\phi_{V}
\end{array}\right),\label{e9}
\end{eqnarray}
 on the ancillary qubit, where $\phi_{V}$ is {
\begin{equation}
\arccos\left(\frac{\sqrt{\left(2s\frac{\sin\frac{\omega t}{2\hbar}}{\omega}\right)^{2}+\cos{}^{2}\frac{\omega t}{2\hbar}}}{\sqrt{\left(2s\frac{\sin\frac{\omega t}{2\hbar}}{\omega}\right)^{2}+\cos{}^{2}\frac{\omega t}{2\hbar}+\left(2s\,\sin\alpha\frac{\sin\frac{\omega t}{2\hbar}}{\omega}\right)^{2}}}\right).\label{e10}
\end{equation}
 } Next, we apply two controlled unitary operations,
\begin{eqnarray*}
C{}_{0-U_{1}}=\left(\begin{array}{cc}
U_{1} & 0\\
0 & I
\end{array}\right),\;\; C{}_{1-U_{2}}=\left(\begin{array}{cc}
I & 0\\
0 & U_{2}
\end{array}\right),
\end{eqnarray*}
 where
\begin{eqnarray}
U_{1} & = & \left(\begin{array}{cc}
\cos\phi_{U_{1}} & i\,\sin\phi_{U_{1}}\\
i\,\sin\phi_{U_{1}} & \cos\phi_{U_{1}}
\end{array}\right),\label{e11}\\
\phi_{U_{1}} & = & \arcsin\left(-\frac{2s\frac{\sin\frac{\omega t}{2\hbar}}{\omega}}{\sqrt{\left(2s\frac{\sin\frac{\omega t}{2\hbar}}{\omega}\right)^{2}+\cos{}^{2}\frac{\omega t}{2\hbar}}}\right),\label{e12}
\end{eqnarray}
 and $U_{2}=\sigma_{z}$. Finally, a Hadamard operation is utilized
on the work qubit $e$. Here, $t$ is the evolution time in the PT-symmetric
system, $s$ is a parameter in the Hamiltonian in Eq. (\ref{e2}),
and $\omega$ is the difference between the two eigenvalues of the
Hamiltonian.

After performing operations shown in Fig. \ref{f1}, the final state
becomes
\begin{equation}
\frac{q}{\sqrt{2}}\left[|0\rangle_{a}e^{-\frac{i}{\hbar}Ht}|0\rangle_{e}+|1\rangle_{a}\frac{1}{q}(\cos\phi_{V}U_{1}-\sin\phi_{V}U_{2})|0\rangle_{e}\right],\label{e13}
\end{equation}
 where $q$ is {
\begin{equation}
e^{\frac{i}{\hbar}ts\cos\alpha}/\sqrt{\left(2s\frac{\sin\frac{\omega t}{2\hbar}}{\omega}\right)^{2}+\cos{}^{2}\frac{\omega t}{2\hbar}+\left(2s\,\sin\alpha\frac{\sin\frac{\omega t}{2\hbar}}{\omega}\right)^{2}},\label{e14}
\end{equation}
 }which is a non-zero number and tends to $1/\sqrt{3}$ as $\alpha\rightarrow-{\pi}/{2}$
and $t=\tau$. If we observe the work qubit conditioned on the ancillary
qubit to be in state $|0\rangle_{a}$, the evolution associated with
the work qubit becomes
\begin{equation}
e^{-\frac{i}{\hbar}Ht}|0\rangle_{e},\label{e15}
\end{equation}
 which is the PT-symmetric Hamiltonian evolution.

 It is worth explaining the symbols. We use $t$ to denote the evolution time in the PT-symmetric quantum system.
 The time it takes to complete the evolution in the work-ancilla two qubits system is designated as $\tilde{t}$. The time it takes for the PT-symmetric quantum system to evolve from $|0\rangle$ to $|1\rangle$ is indicated as $\tau$ and the corresponding time  it takes to complete the operations in the work-ancilla two qubits system represented by $\tilde{\tau}$.
As we construct the PT-symmetric
system with a one-qubit-subspace of a two-qubit-Hilbert space, in which
the sub-system evolves non-unitarily while the whole system evolves
unitarily, the evolving time $\tilde{t}$ of the whole system depends on the evolution time $t$
of the PT-symmetric system, and vice versa. The evolving time $\tau$
for the PT-symmetric system can approach zero when $\alpha$
tending to $-\frac{\pi}{2}$, which is faster than the counterpart
Hermitian system.

\begin{center}
\textbf{\large 4. Experimental Realization}{\large{} }
\par\end{center}{\large \par}

We simulated the evolution process in an NMR quantum system, C$^{13}$-labeled
chloroform that consists of two qubits. The C$^{13}$ nuclear spin
works as the work qubit and the proton nuclear spin works as the ancillary
qubit. We begin from the state $|0\rangle_{a}|0\rangle_{e}$. We then
evolve the PT-symmetric system to some time $t$ by applying the corresponding
operations given in Fig. \ref{f1}. Next, we measure the state of
the work qubit conditioned on the ancillary qubit $a$ being in state
$|0\rangle_{a}$. By varying the instant $t$, we observe the state
evolution of the PT-symmetric system.

The following NMR notations are adopted. The free evolution of the
two-qubit system for a period of $X$ is denoted as $[X]$:
\begin{eqnarray}
\left[X\right]=e^{-i\frac{\pi JX}{2}\sigma_{z}^{a}\sigma_{z}^{e}},
\end{eqnarray}
 where $J=215.23$ Hz is the coupling constant between C$^{13}$ and
H$^{1}$. A rotation of spin $m$ through angle $\phi$ about axis
$j$ is denoted as $[\phi]_{j}^{m}$, and $[\phi]_{j}^{m}=e^{-i\phi\sigma_{j}^{m}/2}$.

The spatial-averaging method \cite{spatialnmr} was used to prepare
the pseudo-pure state $|0\rangle_{a}|0\rangle_{e}$. The single-qubit
rotation operation $V$ is realized by a pulse
\begin{eqnarray}
[2\phi_{V}]_{y}^{a}\label{pulsev}
\end{eqnarray}
 on the ancillary qubit. Here, $C{}_{0-U_{1}}$ and $C{}_{1-U_{2}}$
are realized by the following two pulse sequences:
\begin{eqnarray}
[\frac{\pi}{2}]_{y}^{e}\rightarrow[\frac{\phi_{U_{1}}}{2\pi J}]\rightarrow[\pi]_{x}^{a,e}\rightarrow[\frac{\phi_{U_{1}}}{2\pi J}]\nonumber \\
\rightarrow[\pi]_{-x}^{a,e}\rightarrow[\frac{\pi}{2}]_{-y}^{e}\rightarrow[\phi_{U_{1}}]_{-x}^{e},\label{pulseu1}
\end{eqnarray}
 and
\begin{eqnarray*}
[\pi]_{y}^{e}\rightarrow[\frac{1}{4J}]\rightarrow[\pi]_{x}^{a,e}\rightarrow[\frac{1}{4J}]\rightarrow[\pi]_{-x}^{a,e}\rightarrow[\frac{\pi}{2}]_{-y}^{e}
\end{eqnarray*}

\begin{eqnarray}
\rightarrow[\frac{\pi}{2}]_{x}^{e}\rightarrow[\frac{\pi}{2}]_{-y}^{e}\rightarrow[\frac{\pi}{2}]_{y}^{a}\rightarrow[\frac{\pi}{2}]_{x}^{a}\rightarrow[\frac{\pi}{2}]_{-y}^{a},\label{pulseu2}
\end{eqnarray}
 respectively. Finally, the pulse sequence
\begin{eqnarray}
[\frac{\pi}{2}]_{y}^{a}\rightarrow[\pi]_{-x}^{a},\label{pulseh}
\end{eqnarray}
 is applied to implement the Hadamard operation on the ancillary qubit.
The evolution is observed by looking at both the state of the ancillary
and the work qubit. If the ancillary qubit is in state $|0\rangle_{a}$,
the state of the work qubit gives the evolution under the PT-symmetric
Hamiltonian.

\begin{figure}
\begin{center}
\includegraphics[scale=0.5]{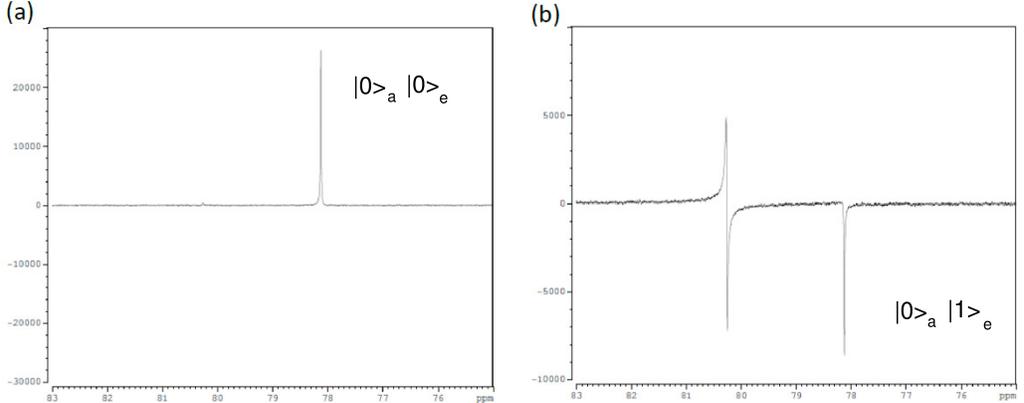}
\caption{Typical spectra of the work qubit with $\alpha=-\pi/8$: (a) pseudo-pure
state at the beginning of evolution; (b) final state after evolving
for a time of $\tau$. \label{f2}}
\end{center}
\end{figure}

\begin{figure}
\centering{}\includegraphics[scale=0.7]{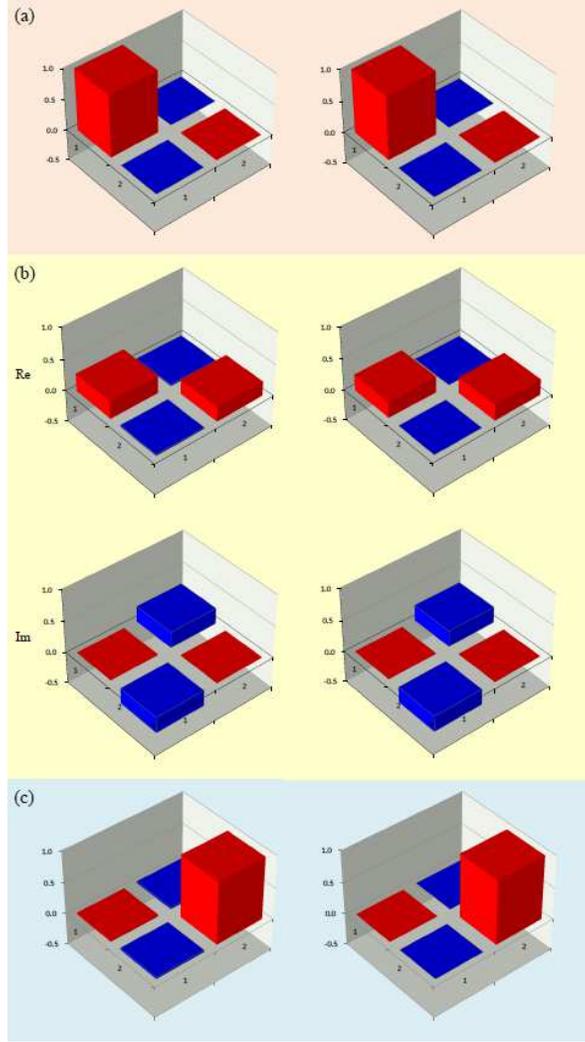} \caption{State tomography of the work qubit with $\alpha=-31\pi/64=-0.4844\pi$:
(a) pseudo-pure state $|0\rangle_{e}$, (b) middle state at $t=\tau/2$,
and (c) final state. In each picture, the left side provides the experimental
results, whereas the right side provides the theoretical results.
(Online version in colour.)\label{f3}}
\end{figure}

Because we are interested in the behavior of the system near $\alpha=-\pi/2$,
we restricted the parameter in the range of $\alpha\in(-\frac{\pi}{2},0]$.
We performed a series of experiments with values of $\alpha$ at 0,
$-\frac{\pi}{4}$, $-\frac{3\pi}{8}$, $-\frac{7\pi}{16}$, $-\frac{15\pi}{32}$,
and $-\frac{31\pi}{64}$. Fig. \ref{f2}(a) shows the spectrum of
the work qubit in the pseudo-pure state. A right single upward peak
in the spectrum indicates that the work qubit and the ancillary qubit
are all in state $|0\rangle$. Fig. \ref{f2}(b) shows the spectrum
of the work qubit for $\alpha=-\frac{\pi}{8}$. The downward peak
on the right indicates that the two-qubit state is $|0\rangle_{a}|1\rangle_{e}$,
whereas the peak on the left corresponds to the ancillary qubit in
state $|1\rangle_{a}$, which is not of interest to the present study.

We now examine the simulation when $t=0$. Here, $V$ operation becomes
an identity operator and it does not require time to simulate. The
two free evolutions $\left[\phi_{U1}/(2\pi J)\right]$ pulses and
the last single qubit pulse $\left[-\phi_{U1}\right]_{x}^{e}$ in
the $C_{0-U_{1}}$ in Eq. (\ref{pulseu1}) also do not require any
time. The other pulses in $C_{0-U_{1}}$ and the whole pulse sequence
of $C_{1-U_{2}}$ and $H$ still require a constant (with respect
to $t$) time to complete. Thus, the simulation performed in the two-qubit
system still requires time $\tilde{t}$ to complete, even though $t$
is zero.

Quantitative results for the evolution in the PT-symmetric system
were obtained by performing quantum state tomography. Fig. \ref{f3}
gives the density matrices of work qubit for $\alpha=-0.4844\pi$
at the beginning ($t=0$), middle ($t=\tau/2$), and end ($t=\tau$)
of the evolution. Fig. \ref{f3}(a) is the density matrix at the beginning
where the state is $|0\rangle$; Fig. \ref{f3}(b) shows the state
in the middle of the process; and Fig. \ref{f3}(c) gives the final
state. For comparison, we drew the corresponding theoretical density
matrices on the right side of each picture. Clearly, the experiments
agree with theory very well.

\begin{figure}
\centering{}\includegraphics[scale=0.8]{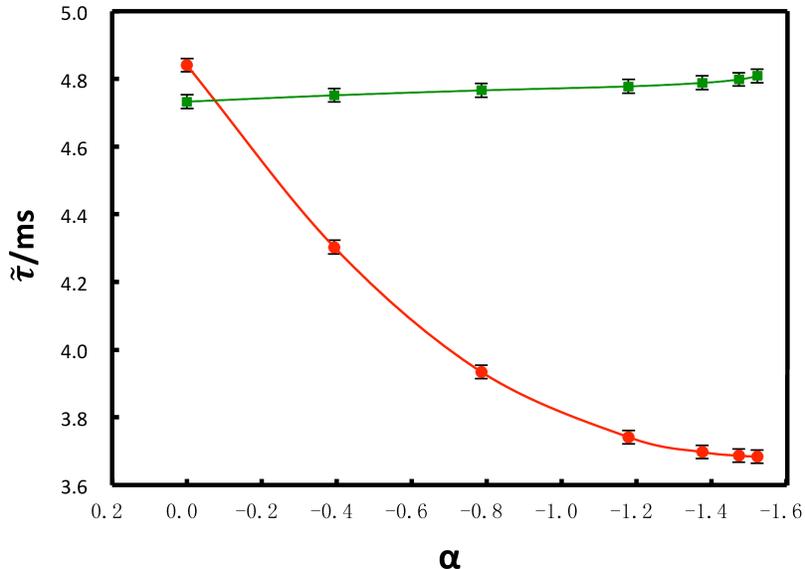} \caption{Total experimental time $\tilde{\tau}$ for simulating evolution from
$|0\rangle$ to $|1\rangle$. The red circles denote the PT-symmetric
case, while the green squares indicate the Hermitian case. The connected
lines are used as visual guides only. (Online version in colour.)\label{f4}}
\end{figure}

The total experimental time $\tilde{\tau}$ to finish the simulation
of the evolution from $|0\rangle_{e}$ to $|1\rangle_{e}$ in the
two-qubit system is shown in Fig. \ref{f4}. As $\alpha$ approaches
$-\pi/2$, $\tilde{\tau}$ decreases remarkably; however, it does
not reach zero. As the PT-symmetric Hamiltonian is realized in a larger
quantum system, $\tilde{\tau}$ depends, not only on the evolution
time $\tau$ in the PT-symmetric system, but also on the time it takes
to set up the PT-symmetric system. The relation between $t$ in the
PT-symmetric system and $\tilde{t}$ in the two-qubit system is determined
through the four operations shown in Fig. \ref{f1}. Of the four operations,
$V$ and $C_{0-U1}$ are dependent on $t$, whereas $C_{1-U2}$ and
$H$ are constant operations that are independent of $t$. In the
simulation, $t$ represents a parameter in determining the operations
of $V$ and $C_{0-U1}$.

\begin{figure}
\centering{}\includegraphics[scale=0.8]{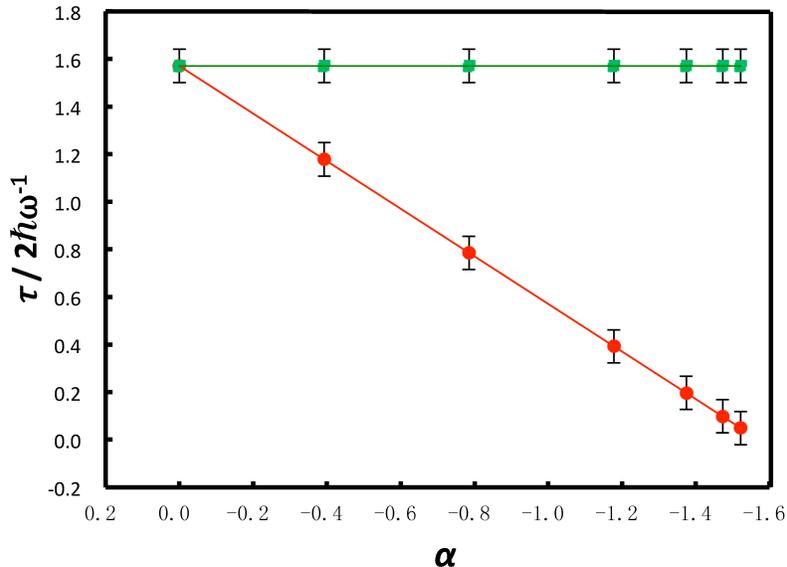} \caption{Evolution time $\tau$ versus $\alpha$. The red circles indicate
the PT-symmetric case, while the green squares denote the Hermitian
case. (Online version in colour.)\label{f5}}
\end{figure}

The relation between $\tau$ and $\alpha$ 
predicted in Ref. \cite{ptbrac} appears in the data remarkably well. An evolution faster than the Hermitian
Hamiltonian system evolution is clearly observed. The evolution time
$\tau$ taken by the PT-symmetric system to go from $|0\rangle$ to
state $|1\rangle$ is shown in Fig. \ref{f5}. In Fig. \ref{f5},
the evolutionary time for the PT-symmetric Hamiltonian system approaches
zero when $\alpha$ approaches $-\pi/2$.

For comparison, we simulated the evolution of the equivalent Hermitian
Hamiltonian ${H_{0}}$. The quantum circuit is similar to that for
the PT-symmetric case shown in Fig. \ref{f1}. However, we substituted
$U_{1}$, $U_{2}$ and $H$ by
\begin{equation}
\tilde{U}_{1}=\tilde{U}_{2}=\left(\begin{array}{cc}
\cos\left(\frac{s\,\cos\alpha}{\hbar}{t}\right) & -i\,\sin\left(\frac{s\,\cos\alpha}{\hbar}{t}\right)\\
-i\,\sin\left(\frac{s\,\cos\alpha}{\hbar}{t}\right) & \cos\left(\frac{s\,\cos\alpha}{\hbar}{t}\right)
\end{array}\right)\label{e17}
\end{equation}
 and $V^{\dagger}$, respectively, where $V^{\dagger}$ is the Hermitian
conjugate of the $V$ operator in Eq. (\ref{e9}). The total time
$\tilde{\tau}$ to implement the simulation and the evolution time
$\tau$ for the equivalent Hermitian system are obtained and shown
in Figs. \ref{f4} and \ref{f5}, respectively. The evolutionary time
$\tau$ for the Hermitian case is clearly constant regardless of the
value of $\alpha$. The faster-than-Hermitian evolution of the PT-symmetric
system is evident.

\begin{center}
\textbf{\large 5. Conclusions}{\large{} }
\par\end{center}{\large \par}

\begin{flushleft}
In conclusion, we experimentally simulated the evolution of a PT-symmetric
system in an NMR quantum system with two qubits. The faster than Hermitian quantum mechanics
evolution of a PT-symmetric system predicted in Ref. \cite{ptbrac}
is clearly observed. When the parameter $\alpha$ approaches $-\pi/2$,
the evolution time also approaches zero.
\par\end{flushleft}

When the difference between the large and small eigenvalues of a Hermitian two-level quantum system is fixed, the fastest evolving time is invariant for a spin flipping in Hermitian quantum mechanics.  However, for PT-symmetric quantum system, the brachistochrone time can be varied by changing the parameters in the Hamiltonian. It can not only accelerate the evolution as predicted in Ref. \cite{ptbrac} and demonstrated in this work, but also decelerate the evolving time as shown in Ref. \cite{decelerated evolution}.


\begin{flushleft}
{\small This work was supported by the National Natural Science Foundation
of China (Grant No. 11175094) and the National Basic Research Program
of China (2009CB929402, 2011CB9216002).}
\par\end{flushleft}


\end{document}